\documentclass{article}

\usepackage{microtype}
\usepackage{graphicx}
\usepackage{booktabs} 

\usepackage{hyperref}

\usepackage[accepted]{icml2023}

\usepackage{amsmath}
\usepackage{amssymb}
\usepackage{mathtools}
\usepackage{amsthm}

\usepackage[capitalize,noabbrev]{cleveref}

\newcommand{\yixuan}[1]{{#1}}

\newtheorem{theorem}{Theorem}[section]
\newtheorem{proposition}[theorem]{Proposition}
\newtheorem{lemma}[theorem]{Lemma}

\theoremstyle{definition}
\newtheorem{definition}[theorem]{Definition}
\newtheorem{assumption}[theorem]{Assumption}
\theoremstyle{remark}
\newtheorem{remark}[theorem]{Remark}

\usepackage{mwe}

\usepackage{subfig}
\usepackage{graphicx}
\usepackage{wrapfig}
\usepackage{multirow}
\usepackage{physics}

\usepackage[textsize=tiny]{todonotes}

\icmltitlerunning{Enforcing Hard Constraints with Soft Barriers: Safe Reinforcement Learning in Unknown Stochastic Environments}

\begin{document}

\twocolumn[
\icmltitle{Enforcing Hard Constraints with Soft Barriers: \\ Safe Reinforcement Learning in Unknown Stochastic Environments}




\begin{icmlauthorlist}
\icmlauthor{Yixuan Wang}{NU}
\icmlauthor{Simon Sinong Zhan}{UCB}
\icmlauthor{Ruochen Jiao}{NU}
\icmlauthor{Zhilu Wang}{NU}
\icmlauthor{Wanxin Jin}{UPenn}
\icmlauthor{Zhuoran Yang}{Yale}
\icmlauthor{Zhaoran Wang}{NU}
\icmlauthor{Chao Huang}{ULiverpool}
\icmlauthor{Qi Zhu}{NU}
\end{icmlauthorlist}

\icmlaffiliation{NU}{Northwestern University, USA}
\icmlaffiliation{UCB}{University of California, Berkeley, USA}
\icmlaffiliation{UPenn}{University of Pennsylvania, USA}
\icmlaffiliation{Yale}{Yale University, USA}
\icmlaffiliation{ULiverpool}{ University of Liverpool, UK}

\icmlcorrespondingauthor{Yixuan Wang}{wangyixu14@gmail.com}

\icmlkeywords{Machine Learning, ICML}

\vskip 0.3in
]



\printAffiliationsAndNotice{}  

\begin{abstract}
Reinforcement Learning (RL) has long grappled with the issue of ensuring agent safety in unpredictable and stochastic environments, particularly
under hard constraints that require the system state not to reach unsafe regions. Conventional safe RL methods such as those based on the Constrained Markov Decision Process (CMDP) paradigm formulate safety violations in a cost function and try to constrain the expectation of cumulative cost under a threshold. However, it is often difficult to effectively capture and enforce hard reachability-based safety constraints indirectly with such constraints on safety violation cost. In this work, we leverage the notion of barrier function to explicitly encode the hard safety chance constraints, and as the environment is unknown, relax them to our design of \emph{generative-model-based soft barrier functions}.  Based on such soft barriers, we propose a novel safe RL approach with bi-level optimization that can jointly learn the unknown environment and optimize the control policy, while effectively avoiding the unsafe region with safety probability optimization. 
Experiments on a set of examples demonstrate that our approach can effectively enforce hard safety chance constraints and significantly outperform CMDP-based baseline methods in system safe rates measured via simulations. 
\end{abstract}

\section{Introduction}\label{sec:intro}
Reinforcement learning (RL) has shown promising successes in learning complex policies for games~\citep{silver2018general}, robots~\citep{zhao2020sim, yang2023tacgnn}, and cyber-physical systems like smart buildings~\cite{Wei_DAC17, Xu_BuildSys21, Xu_BuildSys22},  by maximizing a cumulative reward objective as the optimization goal. However, real-world safety-critical applications, such as autonomous cars~\citep{liu2022physics, liu2023safety, liu2023speculative, luo2023dynamic}, still hesitate to adopt RL policies due to safety concerns. 
In particular, when the environment is stochastic and unknown~\cite{Zhu_ICCAD20, Zhu_ASPDAC21}, these applications often have \emph{hard safety chance constraints} that require the probability of the system state not reaching certain specified unsafe regions above a threshold, e.g., autonomous cars not deviating into adjacent lanes or UAVs not colliding with trees. It is very challenging to learn a policy via RL that can meet such hard safety chance constraints.


In the literature, the Constrained Markov Decision Process (CMDP)~\citep{altman1999constrained} is a popular paradigm for addressing RL safety. Common CMDP-based methods encode safety constraints through a cost function of safety violations, and reduce the policy search space to where the expectation of cumulative discounted cost is less than a threshold. Various RL algorithms are proposed to adaptively solve CMDP through the primal-dual approach for the Lagrangian problem of CMDP. 
However, it is often hard for CMDP-based methods to enforce reachability-based hard safety chance constraints (i.e., the probability bound of the system state not reaching unsafe regions) with the \emph{indirect} constraints on the expectation of cumulative cost. In particular, while reachability-based safety constraints are defined on the system state at the time point level (i.e., each point on the trajectory,), the CMDP constraints only enforce the cumulative behavior in expectation at the trajectory level. In other words, the cost penalty on the system visiting the unsafe regions at a certain time point may be offset by the low cost at other times. 
There is a recent CMDP approach addressing hard safety constraints by using the indicator function for encoding failure probability~\citep{wagener2021safe},  but it requires a safe backup policy for intervention, which is difficult to achieve in unknown environments. Safe exploration with hard safety constraints has also been studied in~\citep{wachi2018safe, turchetta2016safe, moldovan2012safe}. However, these works focus on discrete state and action spaces where the hard safety constraints are defined as a set of unsafe state-action pairs that should not be visited, different from the continuous control setting we are considering.

On the other hand, current control-theoretical approaches for model-based safe RL often try to leverage formal methods to handle hard safety constraints, e.g., by establishing safety guarantees through barrier functions or control barrier functions~\citep{luo2021learning}, or by shielding mechanisms based on reachability analysis~\cite{Huang_EMSOFT19, Fan_ATVA20, Huang_ATVA22} to check whether the system may enter the unsafe regions within a time horizon~\citep{bastani2021safe, Huang_DAC20, wang2020energy, Wang_DAC21, Wang_EMSOFT21, Wang_DT21, Wang_DAC22}. However, these approaches either require explicit known system models for barrier or shielding construction or an initial safe policy to generate safe trajectory data in a deterministic environment. They cannot be applied to the unknown stochastic environments we are addressing.


\begin{figure}
    \centering
    \includegraphics[width=\linewidth]{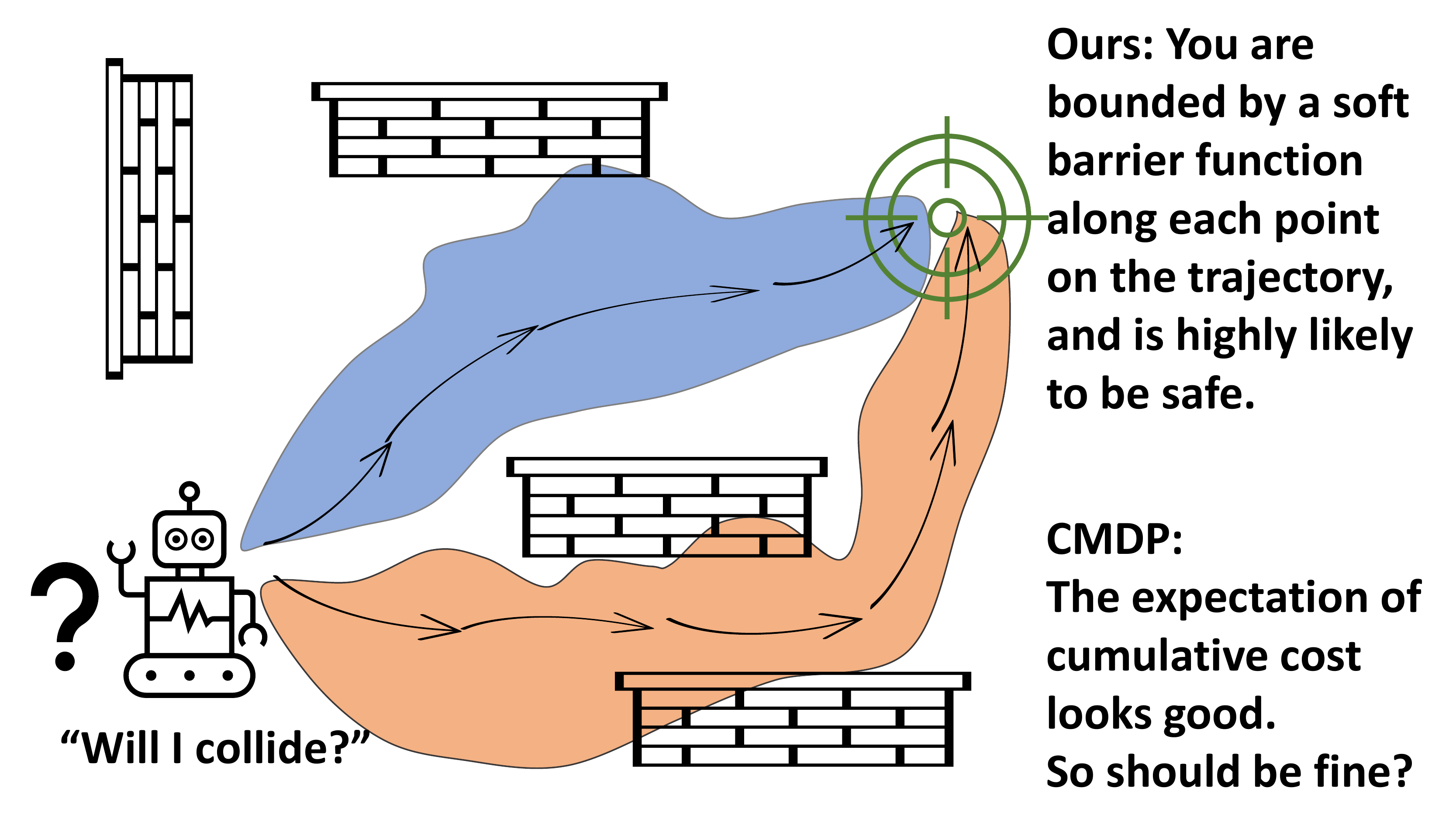}
    \caption{An RL-based robot navigation example that shows the conceptual difference between our approach and CMDP-based ones in encoding the hard safety chance constraints. The satisfaction of CMDP cannot provide any safety probability for the learned policy with any initial state, while our approach can bound/optimize the entire trajectory with a safety probability by the soft barrier function.}
    \label{fig:illustration}
    \vspace{-6pt}
\end{figure}


To overcome the above challenges, we propose a safe RL framework by encoding the hard safety chance constraints via the learning of a \emph{generative-model-based soft barrier function}. Specifically, we formulate and solve a novel bi-level optimization problem to learn the policy with \textbf{joint soft barrier function learning, generative modeling, and policy optimization}. The soft barrier function provides guidance for avoiding unsafe regions based on safety probability analysis and optimization. 
The generative model accesses the trajectory data from the environment-policy closed-loop system with stochastic differential equation (SDE) representation to learn the dynamics and stochasticity of the environment.
And we further optimize the policy by maximizing the total discounted reward of the sampled synthetic trajectories from the generative model. 
This joint training framework is fully differentiable and can be efficiently solved via the gradients. Compared to CMDP-based methods, our approach more directly encodes the hard safety chance constraints along each point of the agent trajectory through the soft barrier function, as shown in Figure~\ref{fig:illustration}. While given the unknown stochastic environment, our approach cannot provide a hard barrier and hence no deterministic safety guarantee, experimental results demonstrate that in simulations, ours can significantly outperform the CMDP-based baselines in system safe rate.



The paper is organized as follows. Section~\ref{sec:related_work} introduces related works, Section~\ref{sec:our_approach} presents our approach, including the bi-level optimization formulation, our safe RL algorithm with generative modeling, soft barrier function learning, and policy optimization to solve the formulation and theoretical analysis of safety probability. Section~\ref{sec:experiments} shows the experiments and Section~\ref{sec:conclusion} concludes the paper.

\section{Related work}\label{sec:related_work}
\textbf{Safe RL by CMDP:} CMDP-based methods encode the safety violation as a cost function and set constraints on the expectation of cumulative discounted total cost~\citep{yang2021wcsac, bharadhwajconservative}. The primal-dual approaches have been widely adopted to solve the Lagrangian problem of constrained policy optimization~\cite{bai2022achieving}, such as PDO~\citep{chow2017risk}, OPDOP~\citep{ding2021provably}, CPPO~\citep{stooke2020responsive},  FOCOPS~\citep{zhang2020first}, CRPO~\citep{xu2021crpo}, and P3O~\citep{shen2022penalized}. \yixuan{Other works leverage a world model learning~\citep{as2021constrained} or the Lyapunov function to solve the CMDP~\citep{chow2018lyapunov}}, or add a safety layer for the safety constraint~\citep{dalal2018safe}.
However, the constraints in CMDP cannot directly encode the hard time-point-level chance constraint, which hinders its application to many safety-critical systems. A recent CMDP-based work uses the indicator function for encoding failure probability as hard safety chance constraints, but it requires a safe backup policy for intervention~\citep{wagener2021safe}.

\textbf{\yixuan{Model-based} Safe RL by Formal Methods:} Formal analysis and verification techniques have been proposed in model-based safe RL to enforce the system not to reach unsafe regions. 
Some works develop shielding mechanisms with a backup policy based on reachability analysis~\citep{shao2021reachability, bastani2021safe}. Other works adopt (control) barrier functions or (control) Lyapunov functions for provable safety~\yixuan{\citep{emam2021safe, choi2020reinforcement, cheng2019end, wang2022joint, ma2021model, luo2021learning, berkenkamp2017safe, taylor2020learning}.}
Moreover, recent work~\citep{yu2022reachability} adopts reachability analysis with CMDP to compute safe feasible sets. However, these methods either require known dynamics, assume a deterministic environment, a safe initial/backup policy, or human intervention, and thus
do not apply to our setting. 


\textbf{Barrier Function for Safety:} Barrier function is introduced as a safety certificate afflicted to the control policy for deterministic and stochastic systems~\citep{prajna2004safety, prajna2004stochastic}. In classical control, finding a barrier function is time-consuming and requires a lot of manual effort, where a common idea is to relax the conditions of the barrier function into optimization formulations such as linear programming~\citep{yang2016linear}, quadratic programming~\citep{ames2016control}, and sum-of-square programming~\citep{wang2022joint}. However, these optimization-based approaches can hardly scale to high-dimensional systems. To this end, recent works have shown great promise in jointly training barrier function and safe policy by neural network representation for better scalability~\citep{qin2021learning}. Our approach leverages the paradigm of barrier function but develops the concept of a soft barrier to address unknown stochastic environments.   

\textbf{RL with Generative Model:} Previous works of generative-model-based RL mainly focus on sample efficiency and policy optimization for the total expected return~\citep{agarwal2020model, li2020breaking, tirinzoni2020sequential}. Some works~\citep{hasanzadezonuzy2021learning, maeda2021reconnaissance} address safe RL by CMDP with a generative model but only solve the tabular discrete state and action space. Besides policy optimization, the generative model in our framework also plays an important role in building a soft barrier function to facilitate the probabilistic safety analysis and optimization. 


\section{Our Approach}\label{sec:our_approach}
\begin{figure}
    \centering    \includegraphics[width=\linewidth]{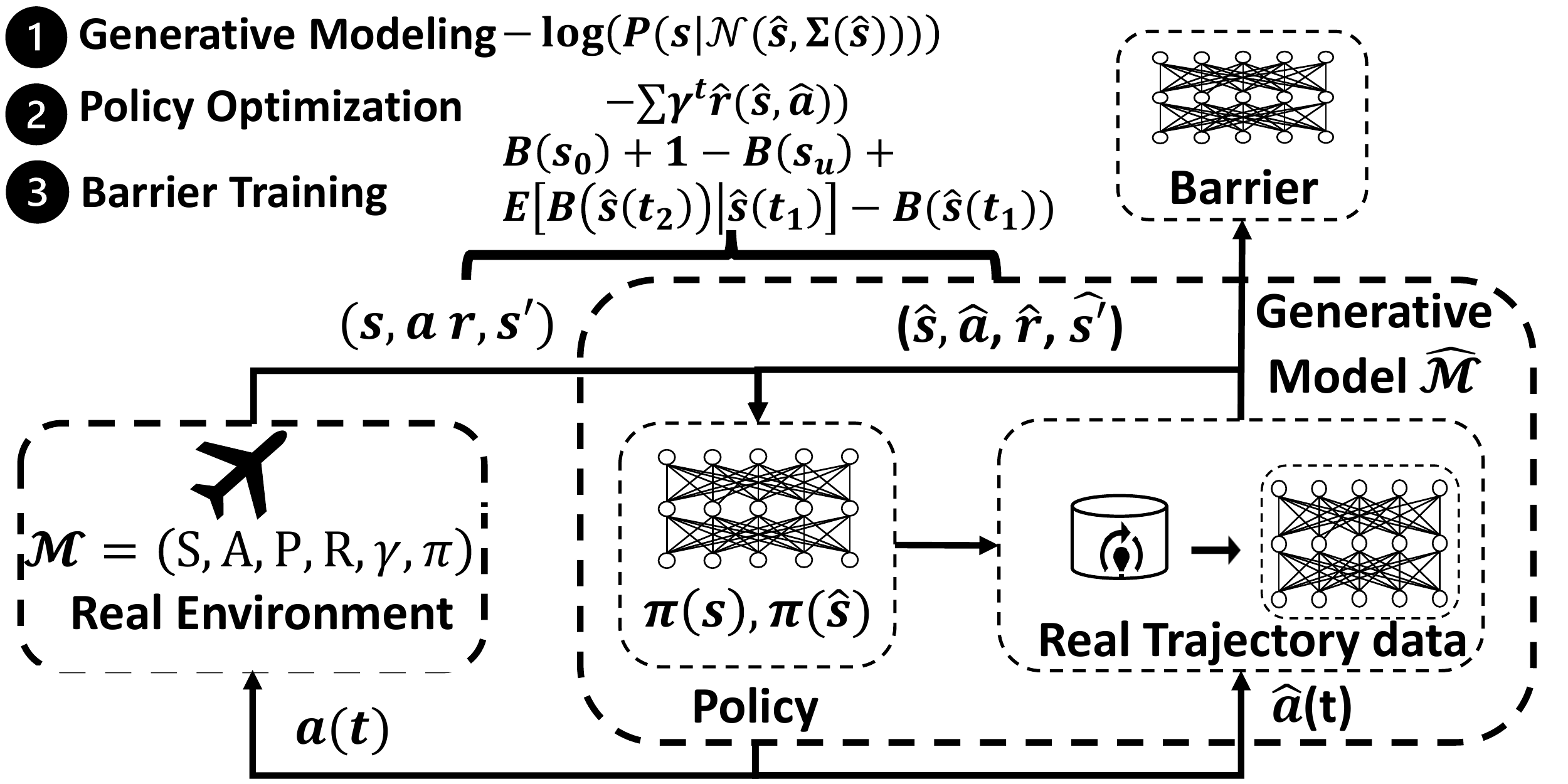}
    \caption{The overview of our safe RL framework based on a generative-model-based soft barrier function. The real environment and generative model share the learning policy and the generative model is abstracted as a discrete-time stochastic differential equation (SDE). We jointly conduct generative modeling, policy optimization, and barrier learning in this framework.}
    \label{fig:overview}
\end{figure}

In this section, we present our framework for safe RL in an unknown stochastic environment that enforces hard safety chance constraints with soft barrier functions. 
In Section~\ref{sec:sub_sec_bilevel_formulation}, we present our bi-level optimization formulation for the problem, which maximizes a total expected return while trying to avoid unsafe regions by optimizing safety probability. Specifically, we encode the hard safety chance constraints with a novel generative-model-based soft barrier function in the lower problem and maximize the performance of the policy with generative model learning in the upper problem. 
We then present our safe RL algorithm to solve the bi-level optimization formulation, by jointly learning the generative model (Section~\ref{sec:subsec-generative-modeling}), soft barrier function (Section~\ref{sec:subsec_soft_barrier}), and  policy optimization (Section~\ref{sec:subsec-policy-optimization}) via first-order gradient, as shown in Figure~\ref{fig:overview}. We conduct theoretical analysis for the safety probability of the learned policy in Section~\ref{sec:subsec-theoretical-analysis}. 

\subsection{Bilevel Optimization Problem Formulation for Safe RL with Soft Barrier}\label{sec:sub_sec_bilevel_formulation}
We assume that the environment can be abstracted as a finite-horizon continuous MDP $\mathcal{M}_{\theta} \sim(\mathcal{S, A, P}, r, \gamma)$, where $\mathcal{S} \subset \mathbb{R}^n$ represents the continuous state space, $\mathcal{A} \subset \mathbb{R}^m$ indicates the continuous action space, and \yixuan{the function class} $\mathcal{P:S \times A \times S \rightarrow } [0, 1] $ denotes the unknown continuous \yixuan{and smooth} stochastic environment dynamics without jump condition. The rewards function $r(s, a) : \mathcal{S} \times \mathcal{A} \rightarrow \mathbb{R}$ is known and the discount factor $\gamma$ $\in [0, 1]$.
A deterministic continuous NN-based policy $\pi_{\theta}: \mathcal{S} \rightarrow \mathcal{A}$ maps the states $s(t) \in \mathcal{S}$ to an action $a(t) \in \mathcal{A}$ at time $t$ as $a(t) = \pi_{\theta}(s(t))$, \yixuan{where $s(t)$ is a random variable at timestep $t$}. 
The environment has several known spaces, i.e., the state space $S \subset \mathcal{S}$,  the initial space $S_0 \subset \mathcal{S}$, and the unsafe space $S_u \subset \mathcal{S}$. The RL objective is to maximize the total discounted expected return as 
\begin{displaymath}
    \max_{\theta} J := \mathbb{E}_{s(0) \in S_0, P(s'|s, a)  }\left[\sum_{\yixuan{t}=0}^{T} \gamma^\yixuan{t} r(s(t), a(t))\right], P \in \mathcal{P}.
\end{displaymath}

\begin{assumption}\label{theorem:assumption}
The dynamics of the environment is assumed to be continuous and smooth. Thus, this paper does not consider discontinuous hybrid dynamics such as contact dynamics in Mujoco and Safety Gym. Such an assumption is not uncommon, as it remains a challenging open problem to learn the discontinuous dynamics~\citep{parmar2021fundamental, pfrommer2021contactnets}. 
\end{assumption}

We address the hard safety chance constraint for RL by requiring the control policy with a safety probability lower bound, as defined below.

\begin{definition}\label{def:safety_probability}
\textbf{(Safety Probability Lower Bound)} A safety probability lower bound $1 - \eta$ of the entire trajectory (process) $\tau_{\theta} = \{s(0), s(1), \cdots, s(T)\}$  is defined as $P\left(s(t) \not\in S_u | s(0) \in S_0, \forall t \in  [0, T]\right) \geq 1 - \eta$, $\eta \in [0, 1]$. 
\end{definition}
\textbf{RL with hard safety chance constraints vs. CMDP:}
Typically, the safe constraints of CMDP are \textit{at the cumulative trajectory cost level as}     
\begin{displaymath}
    \max_{\theta} J(\pi_{\theta})~ \mathrm{s.t.~} \mathbb{E}\left[\sum_{t}\gamma^t c(s(t), a(t))\right] \leq C,
\end{displaymath}
while our safe RL considers more challenging chance constraints (safety probability lower bound) \textit{at the time point level along the entire trajectory as}
\begin{equation}\label{eq:chance_constrained_safe_RL}
\begin{aligned}
    \max_{\theta} &J(\pi_{\theta}),  \\ \mathrm{s.t.~} P\left(s(t) \not \in S_u | \pi_{\theta}, s(0)\right) &\geq 1 - \eta, \forall t \in [0, T], \forall s(0) \in S_0,
\end{aligned}
\end{equation}
where $P$ denotes the safe probability starting from any initial state at any time step. 

\begin{definition} \textbf{(Bi-level Optimization Problem for Safe RL)} To solve the chance-constrained RL in Equation~\eqref{eq:chance_constrained_safe_RL}, we formulate a bi-level optimization problem for our framework as the following, where we use \ $ \hat{}$ \ to denote the elements related to the generative model: 
\begin{align*}
   \max_{\theta, \alpha}  J(\pi_{\theta}) - \lambda \yixuan{\eta^*(\theta, \alpha)^2} - \yixuan{\mathcal{L}_{g} (\tau_{\theta}, \hat{\tau}_{\theta, \alpha})},
\end{align*} 
where $\eta^*(\theta, \alpha)$ denotes the upper bound of unsafe probability  and is the optimal objective to a lower-level problem of the generative-model-based soft barrier function \yixuan{with $\hat{s}$ as the synthetic state in the generative model}: 
\begin{align}\label{eq:bilevel}
& \qquad  \qquad\qquad \qquad \min_{\beta} \eta, \notag  \\
& \mathrm{s.t.}
\begin{cases}
B_{\beta}(\hat{s}) \geq 0, \forall~ \hat{s} \in S, \\
B_{\beta}(\hat{s}) \geq 1, \forall~ \hat{s} \in S_u, \\ 
B_{\beta}(\hat{s}) \leq \eta, \forall~ \hat{s} \in S_0, \\
\mathbb{E}\left[B_{\beta}(\hat{s}(t+1)) | \hat{s}(t) \right] \leq B_{\beta}(\hat{s}(t)), \\ 
\hat{s}(t+1) = \hat{\mathcal{M}}_{\theta, \alpha}(\hat{s}(t)), \forall \ t \in[0, T].
\end{cases}
\end{align}
\end{definition}
Here $\theta$ is the parameter of policy $\pi$. $\alpha$ is the parameter of the generative model $\mathcal{\hat{M}}_{\theta, \alpha} = (\hat{G}_{\alpha}, \hat{\Sigma}_{\alpha})$, which is a stochastic differential equation (SDE) with $\hat{G}_{\alpha}$ as the drift function and $\hat{\Sigma}_{\alpha}$ as the diffusion function for the stochasticity, as shown later in Equation~\eqref{eq:SDE}. $\lambda \geq 0$ is a penalty multiplier. \yixuan{$\tau_{\theta} := \{ s(0), s(1), \cdots, s(T)\}$ and $\hat{\tau}_{\theta, \alpha} := \{ \hat{s}(0), \hat{s}(1), \cdots, \hat{s}(T) \}$ are the sampled realizations of stochastic processes (trajectories) from the environment and from the generative model by the policy $\pi_{\theta}$, respectively.} 
$\beta$ is the parameter of the generative-model-based soft barrier function $B_{\beta}:\mathbb{R}^n \rightarrow \mathbb{R}^+$. 
We encode the hard safety chance constraint by the generative-model-based soft barrier function $B_{\beta}$ in the lower problem, which minimizes \yixuan{$\eta^*(\theta, \alpha)$} as the upper bound of the unsafe probability for $\hat{\mathcal{M}}_{\theta, \alpha}$ in Section~\ref{sec:subsec_soft_barrier}. 
The upper problem aims to optimize the policy's expected return $J(\pi_{\theta})$ and learn the generative model by the \yixuan{maximum likelihood loss  $\yixuan{\mathcal{L}_g(\tau_{\theta}, \hat{\tau}_{\theta, \alpha})}$ between the processes $\tau_{\theta}$ and $\hat{\tau}_{\theta, \alpha}$ as shown later in Equation~\eqref{eq:generative_model_loss}}. Moreover, the upper problem penalizes $\yixuan{\eta^*(\theta, \alpha)}$, which can back propagate the gradient information through $\hat{\mathcal{M}}_{\theta, \alpha}$ to $\pi_{\theta}$ for pushing the agent to avoid $S_u$ in \yixuan{the MDP} $\mathcal{M}_{\theta}$ as long as $\hat{\mathcal{M}}_{\theta, \alpha}$ behaves similar to $\mathcal{M}_{\theta}$. 

We can compute the gradient from $\yixuan{\eta^{*}(\theta, \alpha)}$ for $\pi_{\theta}$ through $\hat{\mathcal{M}}_{\theta, \alpha}$ with current auto-differential tools. This cannot be done in $\mathcal{M}_{\theta}$ as it is unknown. Therefore, the overall bi-level problem is end-to-end differentiable and can be solved efficiently. Figure~\ref{fig:overview} shows how the components in our framework interact with each other. 
The overall algorithm to solve the bi-level problem is shown in Algorithm~\ref{alg:algorithm}. Next, we are going to introduce the details of each module. 

\begin{algorithm}[tb]
\caption{Safe RL with the Generative-model-based Soft Barrier Function}
\label{alg:algorithm}
\textbf{Input}: Unknown environment $\mathcal{M}_{\theta}$ with an initial policy $\pi_{\theta}$ \\
\textbf{Output}: Policy $\pi_{\theta}$ with soft barrier function $B_{\beta}$ based on generative model $\hat{\mathcal{M}}_{\theta, \alpha}$
\begin{algorithmic}[1] 
\FOR {$k$ in $0, \cdots, N$}
\FOR {$i$ in $0, \cdots, M$}
\STATE Sample \yixuan{processes} $\tau_{\theta}^i$ by policy $\pi_{\theta}$  with $\mathcal{M}_{\theta}$ and \yixuan{synthetic processes} $\hat{\tau}_{\theta, \alpha}^i$ by $\pi_{\theta}$ with $\hat{\mathcal{M}}_{\theta, \alpha}$.
\STATE Compute generative loss function $\mathcal{L}_{g}$ with $\tau_{\theta}^i$ and $\hat{\tau}_{\theta, \alpha}^i$ as in Equation~\eqref{eq:generative_model_loss}, $\alpha \leftarrow \alpha - \frac{\partial \mathcal{L}_{g}}{\partial \alpha}.$
\ENDFOR
\STATE Compute barrier function loss $\mathcal{L}_{B}$ by sampling synthetic $\hat{\tau}_{\theta, \alpha}^k$ as in Equation~\eqref{eq:barrier_loss}. 
\STATE Compute total discount reward $\hat{J}(\pi_{\theta})$ by sampling synthetic $\hat{\tau}_{\theta, \alpha}^k$ as in Equation~\eqref{eq:policy_optimization}.
\STATE $\theta \leftarrow \theta - \frac{\partial \mathcal{L}_{B}}{\partial \theta} + \frac{\partial \yixuan{\hat{J}}}{\partial \theta} $ , $\beta \leftarrow \beta - \frac{\partial \mathcal{L}_{B}}{\partial \beta}$.
\ENDFOR
\end{algorithmic}
\end{algorithm}


\subsection{Generative Modeling}
\label{sec:subsec-generative-modeling}

The role of the generative model in our framework is two folds: 
(1) Because the barrier function requires an environment model to encode the hard safety chance constraints, the generative model serves as a surrogate model to build this barrier function, where $\yixuan{\eta^{*}(\theta, \alpha)}$ propagates the gradient to $\pi_{\theta}$ through $\hat{\mathcal{M}}_{\theta, \alpha}$ for improving system safety. (2) The generative model can generate synthetic \yixuan{process (trajectory) $\hat{\tau}_{\theta, \alpha}$} to optimize the performance of the policy efficiently.  

We learn the generative model $\hat{\mathcal{M}}_{\theta, \alpha}$ as a discrete-time SDE to capture the dynamics and stochasticity of the environment and serve as a base for the construction of the soft barrier function:
\begin{equation}\label{eq:SDE}
\hat{\mathcal{M}}_{\theta, \alpha}:\hat{s}(t+1) = \hat{G}_{\alpha}(\hat{s}(t), \pi_{\theta}(\hat{s}(t))) + \hat{\Sigma}_{\alpha}(\hat{s}(t)) W(t),
\end{equation}
where $\hat{G}_{\alpha}: \mathbb{R}^n \times \mathbb{R}^m \rightarrow \mathbb{R}^n$ is an unknown drift function, unknown diffusion function $\hat{\Sigma}_{\alpha}: \mathbb{R}^n \rightarrow \mathbb{R}^{n\times d}$ outputs a $n\times d$ matrix based on $\hat{s}$, and $W(t) \in \mathbb{R}^d$ is the Brownian motion (also known as Wiener Process) with dimension $d$, encoding the stochasticity. When the environment is deterministic, we can simply set the $\hat{\Sigma}(s)$ as $\textbf{0}$. We design the generative model to share the learning control policy with the real environment, as shown in Figure~\ref{fig:overview}. For the inference, the generative model starts from a sample $\hat{s}(0) \in S_0$ and rolls out by drift function $\hat{G}_{\alpha}$, diffusion function $\hat{\Sigma}_{\alpha}$, and policy $\pi_{\theta}$. Therefore, the computation graph contains the learning policy; thus, the auto-differential tools can obtain the gradient for the policy by back-propagating through the generative model. 

\begin{remark}\label{remark:SDE_not_for_hybrid}
We use the fully-connected neural networks to encode such an SDE. Due to the continuity of the neural net, such SDE specification requires the environment dynamics to be continuous and smooth. Therefore our approach cannot handle hybrid dynamics with jump conditions such as the contact dynamics in Mujoco and Safety Gym, as mentioned earlier in Assumption~\ref{theorem:assumption}.
\end{remark}

The generative model training is to reduce the following loss function:
\begin{equation}\label{eq:generative_model_loss}
\begin{aligned}
&\min_{\alpha}\mathcal{L}_g(\tau_{\theta}, \hat{\tau}_{\theta, \alpha}) \\ &= \min_{\alpha} -  \sum_{t=0}^{T}  \log ( P\left(s(t) ~|~ \mathcal{N}(\hat{s}(t), \hat{\Sigma}_{\alpha}(\hat{s}(t)))\right)),
\end{aligned}
\end{equation}
where $\mathcal{L}_g$ is the maximum likelihood loss, $P\left(s(t) ~|~ \mathcal{N}(\hat{s}(t), \hat{\Sigma}_{\alpha}(\hat{s}(t)))\right)$ is the likelihood probability of the observed $s(t)$ under the normal distribution of the SDE representation.    
We use torchsde~\citep{li2020scalable} to fit the data \yixuan{$\tau_{\theta} = \{s(0), s(1), \cdots, s(T)\}$} to the generative model by updating its parameter $\alpha$, which is shown in Lines~2 to~5 in the Algorithm~\ref{alg:algorithm}. 
\subsection{Soft Barrier Function Learning}\label{sec:subsec_soft_barrier}
To encode the hard chance constraint, we introduce a novel generative-model-based soft barrier function.
\begin{definition}
\textbf{(Barrier Function for SDE)} Given a policy $\pi_{\theta}$,  $B_{\beta}$ is a generative-model-based soft barrier function for the discrete-time SDE $\hat{\mathcal{M}}_{\theta, \alpha}$  as in Equation~\eqref{eq:SDE}, if it is twice differentiable and satisfies the constraints of the lower problem in Equation~\eqref{eq:bilevel}.
\label{def:barrier}
\end{definition}
\yixuan{
\begin{lemma}
\cite{prajna2004stochastic} Let $B(\hat{s}(t))$ be a supermartingale of the process $\hat{s}(t)$ and $B(\hat{s}) \geq 0, \forall \hat{s} \in S$. Then for any $\hat{s}(0) \in S_0, c > 0$, $P(\sup_{t\geq 0} B(\hat{s}(t)) \geq c ~|~ \hat{s}(0) \in S_0) \leq \frac{B(\hat{s}(0))}{c}$.
\end{lemma}

\begin{theorem}\label{theorem:barrier_function_safety}
With a barrier function as in Definition~\ref{def:barrier}, the generative-model SDE  with policy $\pi_{\theta}$ (Equation~\eqref{eq:SDE}) has a safety probability lower bound $1 - \eta^*$, \yixuan{where $\eta^*$ is the optimal value in the lower problem of Equation~\eqref{eq:bilevel}}, as $\forall t \in [0, T], P(\hat{s}(t) \not\in S_u | \yixuan{\hat{s}(0) \in S_0}) \geq 1 - \eta^*, \hat{s}(t+1) = \hat{\mathcal{M}}_{\theta, \alpha}(\hat{s}(t)).$
\end{theorem}

\paragraph{Proof:} 
With the last two conditions of the constraints in the lower problem of Equation~\eqref{eq:bilevel}, we have 
\begin{displaymath}
\mathbb{E}\left[B(\hat{s}(t_2)) | \hat{s}(t_1) \right] \leq B(\hat{s}(t_1)), \forall T \geq t_2 \geq t_1 \geq 0,
\end{displaymath}
where $\hat{s}(t_2)$ is the future state of $\hat{s}(t_1)$ by the generative model. This indicates that the barrier function $B(\hat{s})$ is a supermartingale. 
Then by leveraging the Lemma~1 above from~\cite{prajna2004stochastic}, we have 
\begin{displaymath}
\begin{aligned}
&\yixuan{P\left(\hat{s}(t) \in S_u, \textrm{for some } t \in [0, T]~|~ \hat{s}(0) \in S_0 \right)} \\ 
& \leq P\left(B(\hat{s}(t)) \geq 1, \textrm{for some }t \in [0, T]~|~ \hat{s}(0) \in S_0 \right) 
\\ 
&\leq P\left(\sup_{t \in [0, T]}B(\hat{s}(t)) \geq 1 ~|~ \hat{s}(0) \in S_0 \right) \leq B(\hat{s}(0)) \leq \eta^*.
\end{aligned}
\end{displaymath}
Therefore, the safety probability lower bound is $1 - \eta^*$, and Theorem~\ref{theorem:barrier_function_safety} holds. \hfill $\square$}

We further translate the constraints of the lower problem in Equation~\eqref{eq:bilevel} with their sampling mean:  
\begin{displaymath}
    \begin{aligned}
    & \qquad \qquad\qquad \qquad   \min_{\beta} \eta, \notag \\ 
    & \mathrm{s.t.,}\begin{cases}
    \frac{1}{N}\sum_{i=1}^{N} B_{\beta}(\hat{s}^i(0))  \leq \eta, \hat{s}^i(0) \in S_0, \\
    \frac{1}{N}\sum_{i=1}^{N} B_{\beta}(\hat{s}_u^i) \geq 1, \hat{s}_u^i \in S_u, \\
    \frac{1}{N}\sum_{i=1}^{N} B_{\beta}(\hat{s}^i) \geq  0, \hat{s}^i \in S,\\
    \frac{1}{N}\sum_{i=1}^{N} B_{\beta}(\hat{s}^i(t+1)) \leq B_{\beta}(\hat{s}^i(t)), \\  
    \hat{s}^i(t+1) = \hat{\mathcal{M}}_{\theta, \alpha}(\hat{s}^i(t)), t \in [0, T].
    \end{cases}
    \end{aligned}
\end{displaymath}
The third non-negative condition is easy to satisfy by setting the output activation function as \textit{Sigmoid} for the barrier neural network. The last two conditions are to make $B$ as a supermartingale, which is the key to deriving the lower bound of safety probability for the trajectory.  In practice, we use a supervised-learning-based method to optimize this problem by minimizing the following loss function:
\begin{equation}\label{eq:barrier_loss}
\begin{aligned}
\min_{\theta, \beta}&\mathcal{L}_B = \frac{1}{N}\sum_{i=1}^{N} B_{\beta}(\yixuan{\hat{s}^i(0)}) + \frac{1}{N}\sum_{i=1}^{N}(1 - B_{\beta}(\yixuan{\hat{s}^i_u)}) \\ &+ \frac{1}{N}\sum_{i=1}^{N}\left(\frac{1}{M}\sum_{j=1}^{M}B_{\beta}(\hat{s}^{i, j}(t + 1)) - B_{\beta}(\hat{s}^i(t))\right), \\&
\hat{s}^{i, j}(t+1) = \hat{\mathcal{M}}_{\theta, \alpha}(\hat{s}^{i}(t)), t \in [0, T],   
\end{aligned}
\end{equation}
where $\hat{s}^{i, j}(t + 1)$ is the next state of $\hat{s}^{i}(t)$ sampled from the generative model $\hat{\mathcal{M}}_{\theta, \alpha}$ with policy $\pi_{\theta}$. $\mathcal{L}_B$ essentially reduces the barrier mapping value on $S_0$ (the maximum is $\yixuan{\eta^*(\theta, \alpha)}$) and projects the unsafe space $S_u$ to 1 with $Sigmoid$ output, and decreases the expectation of the barrier function along with trajectory. 
It is worth noting that $\mathcal{L}_{B}$ cannot be approximated by the real environment $\mathcal{M}_{\theta}$ with policy $\pi_{\theta}$, as we cannot sample from any intermediate time point $s(t)$ to $s(t + 1)$ to compute the last sample mean in Equation~\eqref{eq:barrier_loss}, which is relatively feasible and simple to do with $\hat{\mathcal{M}}_{\theta, \alpha}$ as in Equation~\eqref{eq:SDE}. The barrier training can be terminated if the second and third sample means in Equation~\eqref{eq:barrier_loss} are zero and non-positive, respectively. The soft barrier training is shown as Line~6 in Algorithm~\ref{alg:algorithm}. 

\subsection{Policy Optimization}
\label{sec:subsec-policy-optimization}

As stated before, we use the generative model to generate synthetic data \yixuan{$\hat{\tau}_{\theta, \alpha}^i = \{\hat{s}^{i}(0), \cdots, \hat{s}^{i}(T)\} (i \in [1, N])$} with policy $\pi_{\theta}$ to maximize the total expected return $\yixuan{\hat{J}(\pi_{\theta})}$ as:
\begin{displaymath}
\begin{aligned}
\max_{\pi_{\theta}} \hat{J}(\pi_{\theta}) = \mathbb{E}_{\hat{s}(0), \mathcal{\hat{M}}_{\theta, \alpha}} \left[ \sum_{t=0}^{T} \gamma^{t} r\left(\hat{s}(t), \pi_{\theta}(\hat{s}(t))\right)\right] \\ \textrm{s.t.}~
\yixuan{\hat{s}(t+1) = \hat{\mathcal{M}}_{\theta, \alpha}(\hat{s}(t)),\forall t \in [0, T]}.
\end{aligned}
\end{displaymath}
We use the sample mean from the synthetic trajectories as an estimate for the expectation:
\begin{equation}\label{eq:policy_optimization}
\begin{aligned}
\max_{\pi_{\theta}}  \hat{J}(\pi_{\theta}) =  \frac{1}{N}\sum_{i=0}^{N}  \sum_{t=0}^{T} \gamma^{t} r\left(\hat{s}^i(t), \pi_{\theta}(\hat{s}^i(t))\right), \\ \textrm{s.t.} \
\yixuan{\hat{s}^i(t+1) = \hat{\mathcal{M}}_{\theta, \alpha}(\hat{s}^i(t)),\forall t \in [0, T]}.
\end{aligned}
\end{equation}
With policy $\pi_{\theta}$ in the computation graph of  $\mathcal{\hat{M}}_{\theta, \alpha}$, we can directly obtain the backwards gradient for $\pi_{\theta}$ from Equation~\eqref{eq:policy_optimization}. The policy optimization is shown as Line~7 in the Algorithm~\ref{alg:algorithm}.

\subsection{Theoretical Analysis of Safety Probability under Soft Barrier}
\label{sec:subsec-theoretical-analysis}

For the \textit{final learned} policy, we conduct a theoretical analysis of its safety probability (as defined in Definition~\ref{def:safety_probability}), derived from the generative-model-based soft barrier function in our framework.

\begin{lemma}\label{lemma:asym_generative_model_bound}
\textbf{(Theorem 21 in~\citep{agarwal2020flambe})} Given $\delta \in (0, 1)$, a learned deterministic policy $\pi_{\theta}(s)$ and assume the environment-policy transition dynamics as $\yixuan{P^*(s' | s)} \in \mathcal{P}$ with the function class $\mathcal{|P|} < \infty$ \yixuan{($s'$ represents the next state of $s$)}, let the environment and policy generate a dataset of $n$ trajectories $D := \{(s^j(t), s^j(t+1))\}_{t=0}^{T} (j = 1, \cdots, n)$, \yixuan{$s(t) \sim D^t = (s^j(0:t-1))$. Note that $D^t$ is a martingale depending on the previous examples.} \yixuan{Let the generative model $\hat{\mathcal{M}}_{\theta, \alpha}$ maximize the likelihood of the dataset by its transition dynamics $\hat{P}$ via Equation~\eqref{eq:generative_model_loss}.}
Then with at least probability $1 - \delta$,  the expectation of total variation distance between $P^*$ and $\hat{P}$ is bounded as:
\begin{equation}
\begin{aligned}
&\sum_{t=0}^{T} \mathbb{E}_{\yixuan{s \sim D^t}}
\left[d_{\textrm{TV}}(P^*, \hat{P})\right] \\ &= \sum_{t=0}^T\mathbb{E}_{\yixuan{s \sim D^t}} \norm{\hat{P}(s'|s) - P^*( s'|s)}_{\mathrm{TV}}^{2} \leq 
\frac{2\log(|\mathcal{\yixuan{P}}| / \delta)}{n}.
\label{eq:generative_bound}
\end{aligned}
\end{equation}
\end{lemma}

\begin{lemma}\label{non_negative_limit}
Given a random variable $X_n \geq 0$ on a probability space $\Omega$, if \ $\mathbb{E}_{\Omega}[X_n] \rightarrow 0$  \ as \ $n \rightarrow \infty$, then $P(X_n = 0) \rightarrow 1$. 
\end{lemma}

\textbf{Proof:}
For any $m \in \mathbb{N}$, let  $E_m = \{\omega \in \Omega: X_n(w) > \frac{1}{m} \}$. Since $X_n \geq 0$, we have:
\begin{displaymath}
\mathbb{E}_{\Omega}[X_n] = \int_{\Omega} X_n \mathrm{d}P \geq \int_{E_m} X_n \mathrm{d}P \geq \frac{1}{m} P(E_m).
\end{displaymath}
Therefore, $P(E_m) \rightarrow 0$, and then:
\begin{displaymath}
\begin{aligned}
0 \leq P(\{\omega \in \Omega: X_n(w) \neq 0\}) &= P(\bigcup E_m) &\\ = \lim_{m\rightarrow \infty} P(E_m) \rightarrow 0, \\
\end{aligned}
\end{displaymath}
\begin{displaymath}
\begin{aligned}
P(\{\omega \in &\Omega : X_n(w) \neq 0\}) \rightarrow 0 \\ &\implies P(\{\omega \in \Omega: X_n(w) = 0\})  \rightarrow 1.
\end{aligned}
\end{displaymath}


\begin{proposition}\label{prop:theoretical}
\textbf{(Asymptotic Lower Bound of Safety Probability)} Given the learned policy $\pi_{\theta}$, let the generative model fit $n$ sample trajectories $\tau^i_{\theta} (i = 1, \cdots, n)$ from environment $\mathcal{M}_{\theta}$ with $\pi_{\theta}$ by Equation~\eqref{eq:generative_model_loss}, learn the generative-model-based soft barrier function $B_{\beta}$ by Equation~\eqref{eq:barrier_loss} with $\eta^*$ and assume that it formally satisfies the constraints in Equation~\eqref{eq:bilevel},  then the real environment $\mathcal{M}_{\theta}$ with policy $\pi_{\theta}$ is safe with probability at least  $(1-\eta^*)$ when $n\rightarrow \infty$.
\end{proposition}

\textbf{Proof of Proposition~\ref{prop:theoretical}}: Given $(S, \mathcal{B})$ as the measure spaces with $S$ as the state space and $\mathcal{B} = \{B: \mathcal{S} \rightarrow \mathbb{R}, \norm{B}_{\infty} \leq 1 \}$, where $B$ is a generative-model-based soft barrier function with $Sigmoid$ output, then according to the definition of total variation distance and Lemma~\ref{lemma:asym_generative_model_bound}, we can bound the expectation of the difference between the barrier values of the real trajectory and the synthetic trajectory as
\begin{displaymath}
\begin{aligned}
&\sum_{t=0}^T\mathbb{E}_{\yixuan{s \sim D^t}}\left[\frac{1}{2}\sup_{B\in\mathcal{B}} \mathbb{E}_{P^*(s' | s)} [ B(s')] - \mathbb{E}_{\hat{P}(s'|s)} [B(s')] \right] \\
& =\sum_{t=0}^{T}\mathbb{E}_{\yixuan{s \sim D^t}}
\left[d_{\textrm{TV}}(P^*, \hat{P})\right] \leq \frac{2\log(|\mathcal{\yixuan{P}}| / \delta)}{n}.
\end{aligned}
\end{displaymath}

When $n \rightarrow \infty$, we set $\delta = \frac{1}{n}$ and let $X_n =  \frac{1}{2}\sup_{B\in\mathcal{B}} \mathbb{E}_{P^*(s' | s)} [ B(s')] - \mathbb{E}_{\hat{P}(s'|s)} [B(s')]$, and therefore $\mathbb{E}[X_n] \rightarrow 0$. We know $X_n \geq 0$, since $X_n = 0$ when $P^* = \hat{P}$. Therefore, according to Lemma~\ref{non_negative_limit}, $P(X_n \rightarrow 0) \rightarrow 1$.  
We then assume $D^t \yixuan{(t \in [0, T])}$ can uniformly cover the space $S$ as $n \rightarrow \infty$, thus the soft barrier becomes a true barrier function for the real environment and Proposition~\ref{prop:theoretical} holds. \hfill $\square$





\smallskip
\begin{remark}\label{practical_lower_bound}
\textbf{(Practical Safety Probability Lower Bound)} In addition to the asymptotic safety probability, we propose a finite-sample practical safety probability lower bound.
We first sample the generative model and the environment with the final learned policy to quantify their maximum distance per state as 
$\Delta = \max_{t \in [0, T], i = 1, \cdots, N}|s^i(t) - \hat{s}^i(t)|$, 
and then enlarge the unsafe region with $\Delta$ by Minkowski sum as
$S_u^{'} = S_u \bigoplus \Delta$. 
Next, we retrain another generative-model-based soft barrier function $B$ with $S_u'$.
Finally, we conservatively report  $(1 - \max_{(\hat{s} \in \hat{\tau}_t^i, i =1, \cdots, N)} B(\hat{s}^i_t))$ as the final lower bound of safety probability by the soft barrier function.
\end{remark}

\smallskip
\begin{remark}
\textbf{(During-learning Safety)} The above asymptotic and practical safety bounds are derived for the final learned policy. It is possible that $1-\eta^*$ is not a valid safety probability bound during learning, as there exists a modeling gap between the generative model and the real environment. However, we optimize $1 - \eta^*$ during learning to increase the chance of finding safer learned policies at the end, as demonstrated in our experiments below.  
\end{remark}


\section{Experimental Results}\label{sec:experiments}
 \textbf{Experiment Settings and Examples:} 
 As stated in Section~\ref{sec:related_work}, other model-based safe RL methods with hard safety constraints either require known dynamics, a
safe initial/backup policy, or human intervention, and thus do not apply to the problem setting we are considering. Therefore, we compare our approach with two state-of-the-art open-source model-free CMDP-based methods, PPO-L~\citep{ray2019benchmarking} and FOCOPS~\citep{zhang2020first}. For these two baselines, we design the cost function such that the state is safe if its cost is less than $0$. It is worth noting that PPO-L has a stronger safety constraint than FOCOPS as we implemented the PPO-L with the expectation of cost per state as $\mathbb{E}[c(s, a)] \leq 0$, rather than the cumulative cost in FOCOPS as $\mathbb{E}\left[\sum_{t=0}^{T} c(s, a)\leq D' \right]$.  In FOCOPS, We conservatively set $D' = -60$ for the 2D and cartpole examples below, and $-200$ for the Rocket and UAV examples, to improve its safety. We mark this safety-oriented version as FOCOPS*. 
We mainly compare the converged final policy of each method in system safe rate measured via simulations -- we call it \emph{empirical safe rate}. We also perform safety probability analysis for our method (CMDP cannot provide one), and compare different methods in total reward return. 

Note that learning safe control policy for high-dimensional systems under hard safety constraints is quite challenging. Current state-of-the-art works of certificate-based policy learning mainly focus on low-dimensional systems with fewer than 9D states~\citep{luo2021learning, lindemann2021learning,chang2019neural, berkenkamp2017safe,  dawson2022safe}. In this paper, among the four examples shown below, we are able to test our approach on 13D UAV and Rocket examples:



\begin{table*}
  \caption{Comparison of our approach with CMDP-based baselines PPO-L and FOCOPS*. $s_e$ is the safe rate by simulating 500 random initial states from $S_0$. $1-\eta$ is the practical lower bound of safety probability in our approach as $(1 - \max_{(\hat{s} \in \hat{\tau}_t^i, i =1, \cdots, n)} B(\hat{s}^i_t))$, derived by \textbf{Remark~\ref{practical_lower_bound}}. We report the mean and std values (in parenthesis) for 5 individual runs. Our approach achieves significantly higher $s_e$ than the baselines. It is observed that $1 - \eta$ is a lower bound of $s_e$. }
  \centering
  \begin{tabular}{l | c | cccl}
    \toprule
    Metric & Methods & 2D & Cartpole & Rocket & UAV\\
    \midrule
    \multirow{3}*{\begin{tabular}{l} $s_e$, empirical\\
     safe rate
    \end{tabular}}
    & Ours & \textbf{99.9(0.09)\%} & \textbf{100\%} & \textbf{100\%} & \textbf{100\%}  \\
    & PPO-L & 98.9(0.08)\% & 89.3(5.5)\% & 96.4(6.3)\% & \textbf{100\%} \\
    & FOCOPS* & 98.7(0.18)\% & 84.2(4)\% & \textbf{100}\% & 91(4.2)\%\\
    \midrule
    \multirow{2}*{\begin{tabular}{l} $1-\eta$,\\
    safety lower bound
    \end{tabular}}
    &Ours & 97.6(1.3)\% & 86.6(2.9)\% & 89.9(1.6)\% & 93.2(2.2)\%  \\ 
    & PPO-L, FOCOPS* & - & - & - & - \\
    \midrule
    \multirow{3}*{\begin{tabular}{l} $J(\pi)$,\\
    performance
    \end{tabular}}
    & Ours & -67.3(4.9) & -24.4(4.7) & \textbf{-143.2}(1.6) & -847.1(6.5) \\
    & PPO-L & \textbf{-66.3}(5.3) & -34.1(6.7) & -151.4(3.6) & -895.5(4.3) \\
    & FOCOPS* & -69.8(3.2) & \textbf{-15.2}(2.3) & -249.1(1.4) & \textbf{-734.1}(3.3) \\
  \bottomrule
\end{tabular}\label{exp:table}
\end{table*}

\textit{\underline{2-Dimensional SDE~\citep{prajna2004stochastic}}} has the unknown dynamics $\mathcal{M}$ as $\dot{s_1} = 0.8s_2, \mathrm{d}{s_2} = (a - 0.3s_1^3)dt + 0.2\mathrm{d}W(t)$ ($W(t)$, Wiener process.) 
Initial space $S_0 = \{(s_1 + 2)^2 + s_2 \leq 0.01  \}$, and unsafe space $S_u = \{s_1 \in [-1, 0], s_2 \in [1.2, 1.7]\}$. The goal is to stabilize the system near $(0, 0)$.

\textit{\underline{Cartpole Balancing}} has a 4-dimensional vector $s = [x, \theta, \dot{x}, \dot{\theta}]$ as the system state, where $x$ is the position and $\theta$ is the angular error to the upright. The initial space $S_0 = \{(x, \theta, \dot{x}, \dot{\theta}) | x \in [-0.167, 0.033], \theta \in [-0.6, -0.5],  \dot{x} = -0.35, \dot{\theta} = 0.53 \}$, and unsafe space $S_u =  \{(x, \theta, \dot{x}, \dot{\theta}) \ | \ x \leq -0.75 \}$. The goal is to keep the cartpole balanced upright.

\textit{\underline{Powered Rocket Landing~\citep{jin2021safe}}} has 6 DoF (degrees of freedom) with 13 system states and 3 action variables. The goal is to land the rocket close to the original point while avoiding an unsafe region. Its state vector is
$\mathbf{s} = \left[\textbf{p \ v \ q} \ \mathbf{\omega} \right] \in \mathbb{R}^{13}$,
where $\textbf{p} = (x, y, z) \in \mathbb{R}^3$ and $\textbf{v} = (v_x, v_y, v_z) \in \mathbb{R}^3 $ represent the position and velocity of the rocket, respectively. $\textbf{q} \in \mathbb{R}^4$ is the unit quaternion for attitude and $\mathbf{\omega} \in \mathbb{R}^3$ is the angular velocity with respect to the inertial frame. There are three trust forces for the rocket as the control input
$\textbf{u} = \left[T_x, T_y, T_z \right] \in \mathbb{R}^3$. 
Initial space $S_0: \textbf{p} = (x, y, z) (x - 10)^2 + (y + 8)^2 + (z - 5)^2 \leq 0.01,   \textbf{v} = 0, \textbf{q} = (0.73, 0, 0, 0.68), \omega = 0$, and unsafe space $S_u: \textbf{p} = (x, y, z) (x - 5)^2 + y^2 \leq 1, -2\leq z \leq 5, \norm{\textbf{v}}_1 \leq 10, \norm{\mathbf{\omega}}_1 \leq 10$.

\textit{\underline{UAV Maneuvering~\citep{jin2021safe}}}
is to maneuver a UAV close to the original point while avoiding an obstacle. The 6-DoF UAV has 13 system states and 4 action variables. Its state vector is 
$\mathbf{s} = \left[\textbf{p \ v \ q} \ \mathbf{\omega} \right] \in \mathbb{R}^{13}$, same with above Rocket example.
 The control input $\mathbf{u} = \left[ T_1, T_2, T_3, T_4 \right] \in \mathbb{R}^4$ includes the four rotating propellers of the quadrotor.
Initial space $S_0: \textbf{p} = (x, y, z) (x + 8)^2 + (y + 6)^2 + (z - 9)^2 \leq 0.01, \textbf{v} = 0, \textbf{q} = (1, 0, 0, 0), \omega = 0 $, unsafe space 
$S_u: \textbf{p} = (x, y, z) (x + 4.5)^2 + (y + 4)^2 \leq 1, -2\leq z \leq 5. $

\textbf{Comparison and Effectiveness of Our Approach:}
Table~\ref{exp:table} shows the comparison results in simulation-based system safe rate (based on 500 simulations for each example, with random initial states), safety probability, and performance for 5 individual runs. We can see that \textbf{by directly enforcing hard safety constraints via soft barrier functions,} \textbf{our approach can achieve significantly higher system safe rates than the CMDP-based baselines.  Our approach also provides a practical lower bound of safety probability, which the CMDP-based methods cannot provide}. CMDP achieves better performance (total reward return) in some cases, but we view safety as the first priority for these systems and the focus of this work. 
\begin{figure}[!ht]
    \subfloat[2D system]{\includegraphics[width=0.5\linewidth,height=0.5\linewidth]{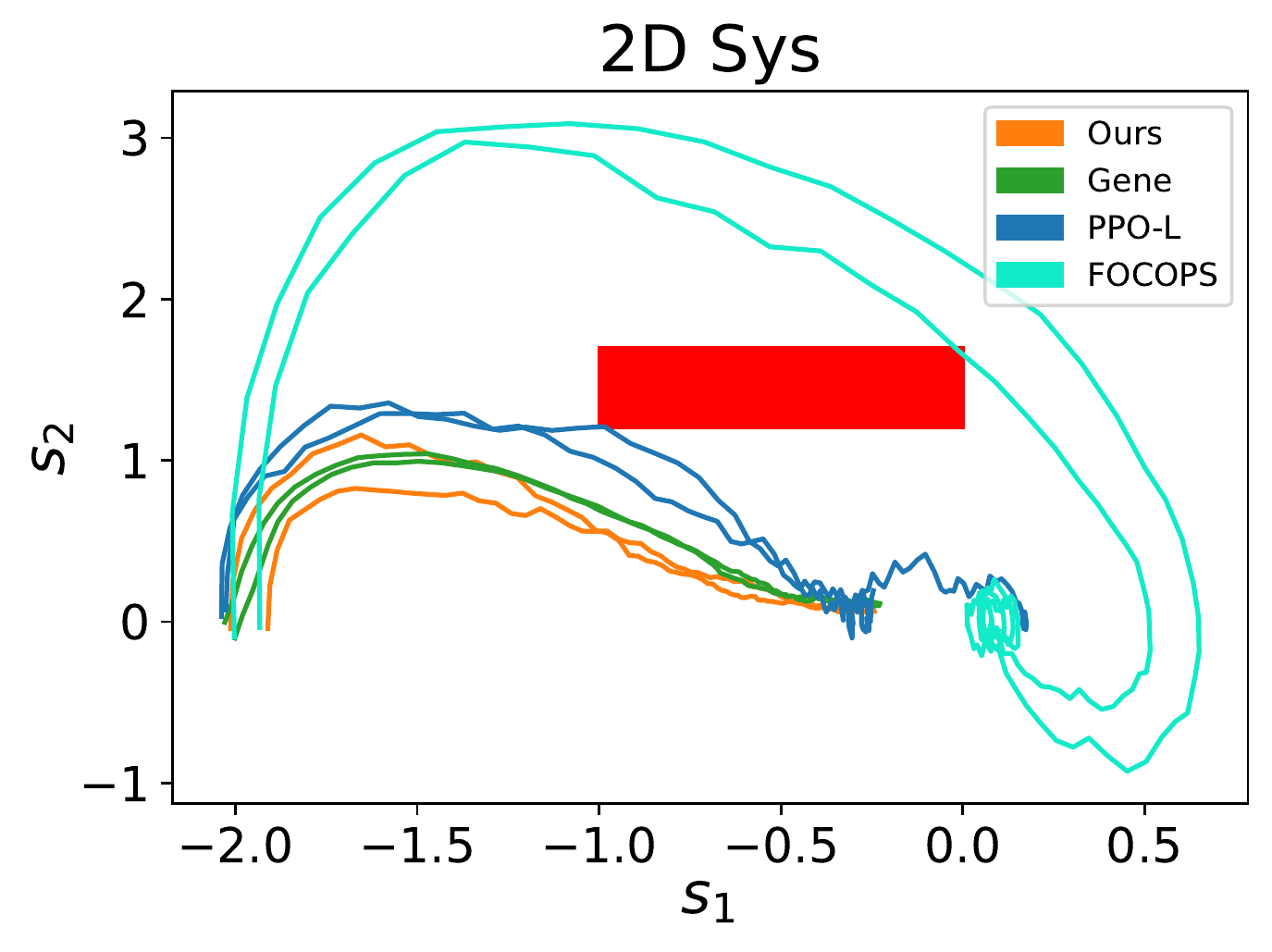}} 
    \subfloat[Cartpole balancing]{\includegraphics[width=0.5\linewidth,height=0.5\linewidth]{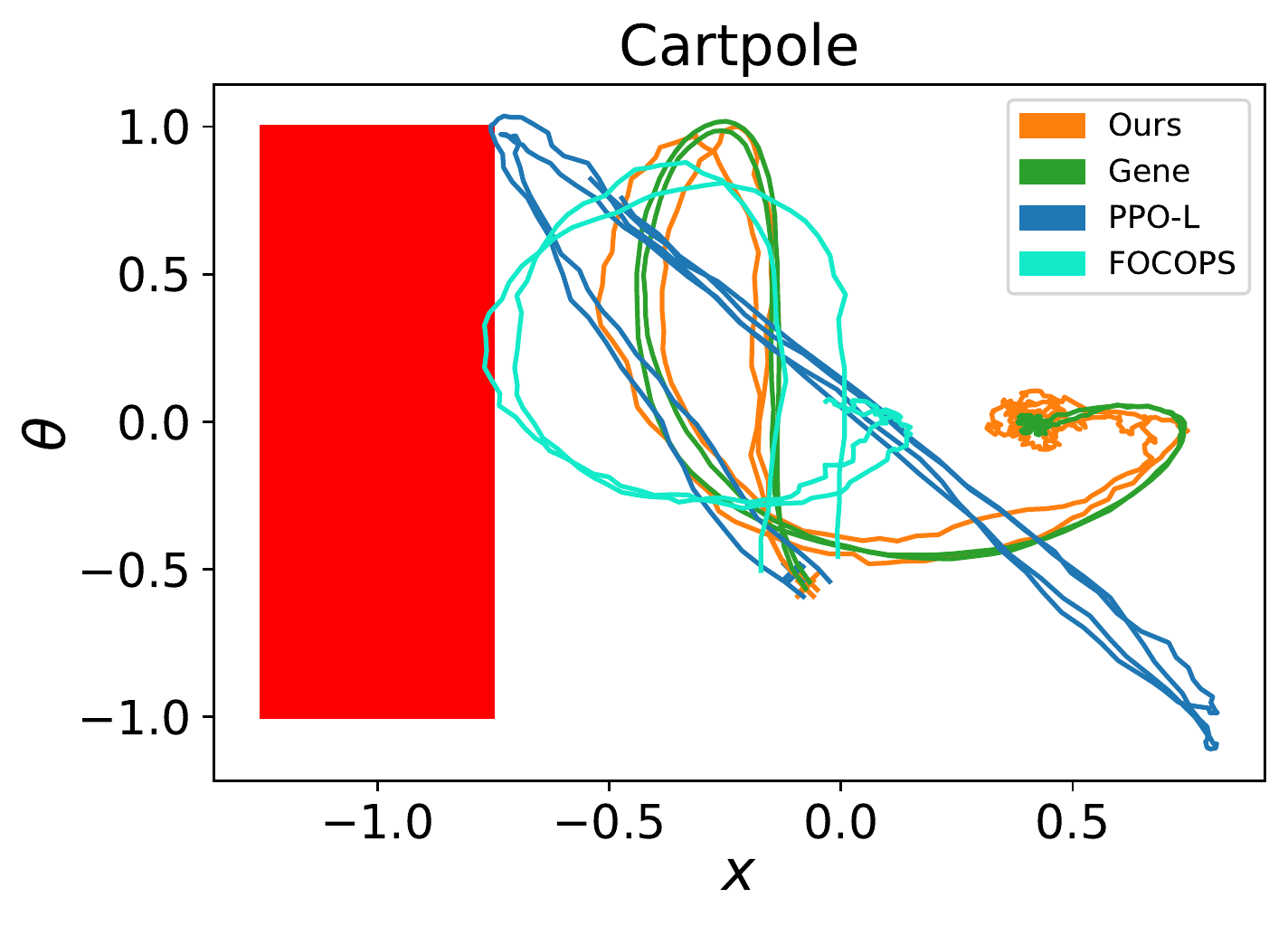}} \\
    \subfloat[Rocket landing]{\includegraphics[width=0.5\linewidth,height=0.5\linewidth]{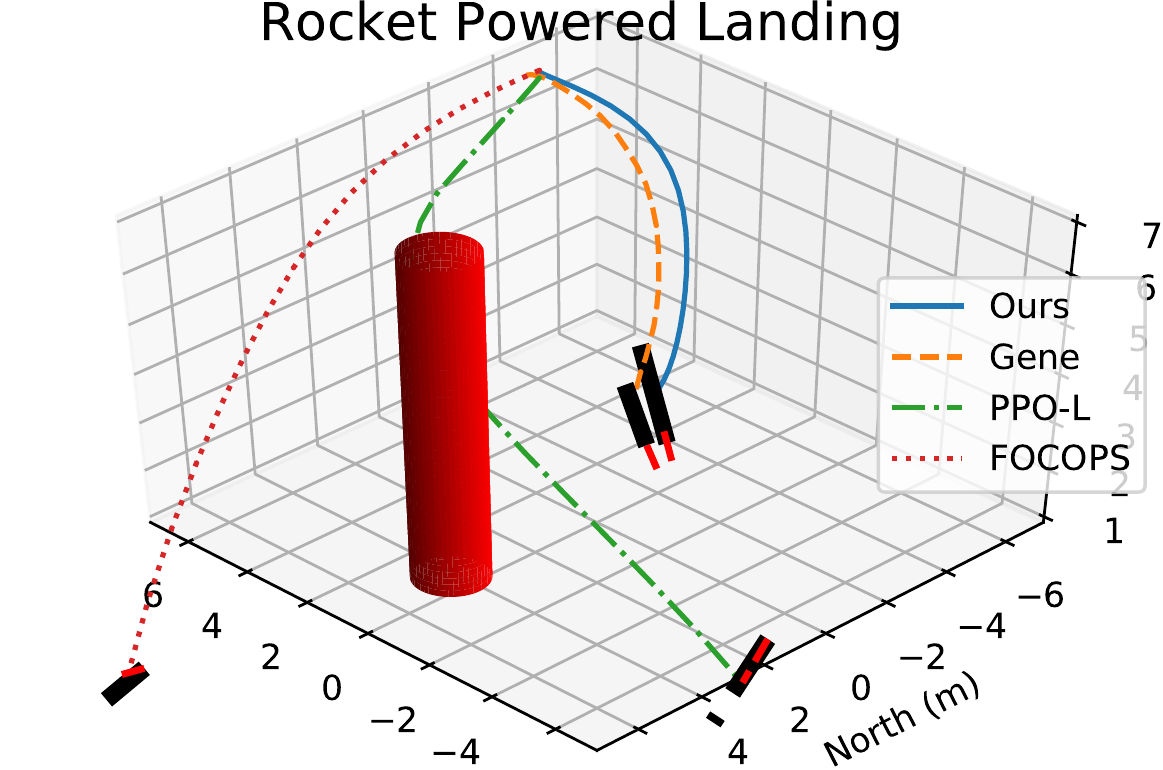}}%
    \subfloat[UAV maneuvering]{\includegraphics[width=0.5\linewidth,height=0.5\linewidth]{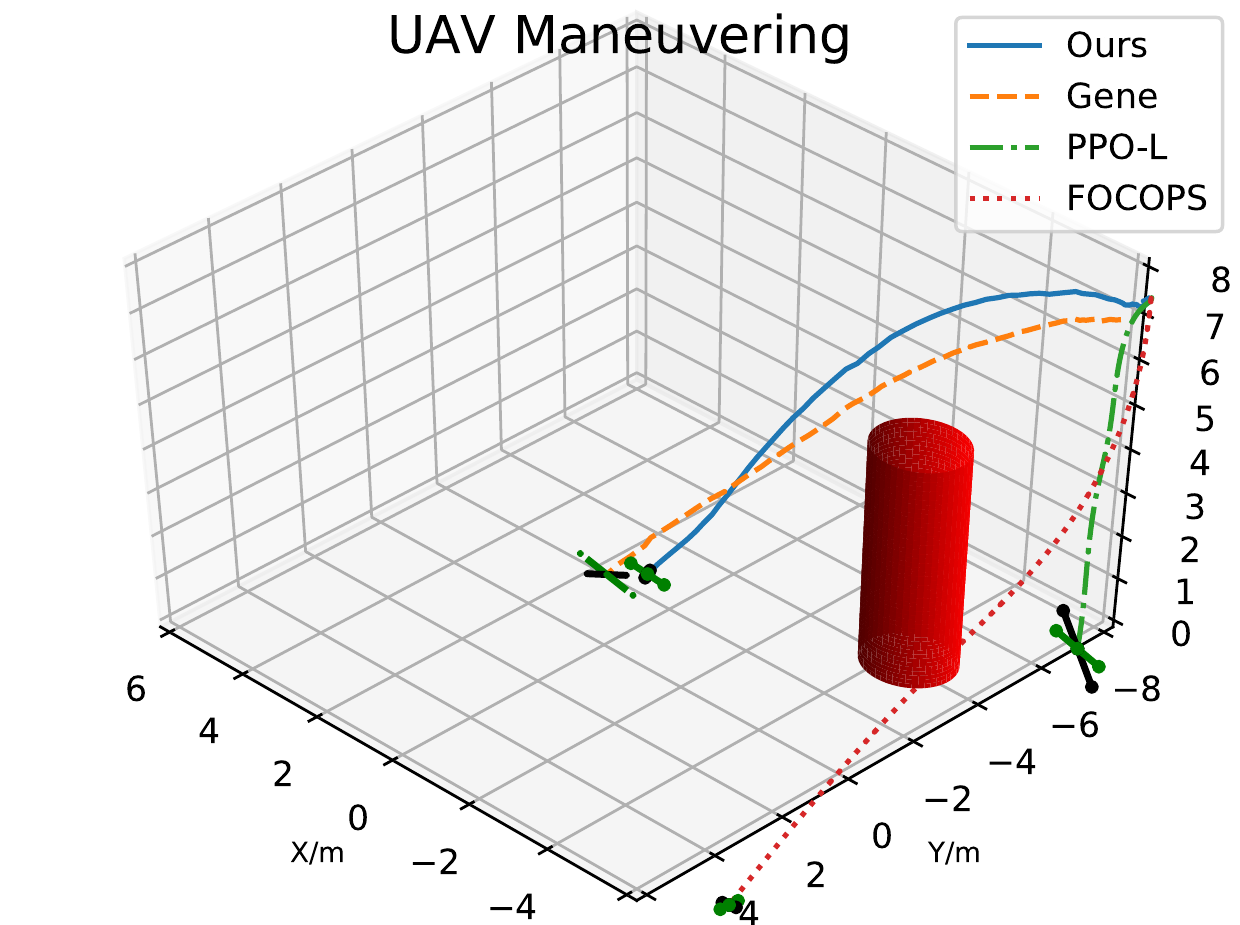}}
    \caption{Control trajectories by the learned policies from our approaches and baselines. \yixuan{``Gene" indicates the synthetic trajectory from the final learned generative model, which behaves very similarly to the real environment with the ``Ours" policy, showing its effectiveness for barrier function construction.} We can see that our approach learns safer policies than the baselines. }
    \label{fig:all_examples}
\end{figure}

Figure~\ref{fig:all_examples} shows the control trajectories by the learned policies from our approach and the baselines. The agent is safe with our learned policy, while there exist unsafe cases by both PPO-L and FOCOPS. Moreover, our generative model behaves very similarly to the real environment, which shows the usefulness of the generative modeling for constructing the soft barrier function and optimizing the control policy.
We also show the learning process of the soft barrier function on the generative model and its testing in the real environment in Figures~4-7 for all the examples. The learned barrier function maps the initial space to near $0$ and the unsafe space to $1$ with the third sample mean in Equation~\eqref{eq:barrier_loss} to $0$ (marked as Lie in Figures). The barrier function has a similar value along with the trajectories in the real environment and the generative model. Again, this indicates that the generative model behaves very similarly to the environment, as shown in Figure~\ref{fig:all_examples}. Although the barrier function decreases or stays constant most of the time, it can increase at some point. 
This is due to 1) the possible modeling error between the generative model SDE and environment and 2) the supervised learning approach cannot cover all possible cases for soft barrier function training.


\begin{figure}
    \centering
    \includegraphics[width=\linewidth]{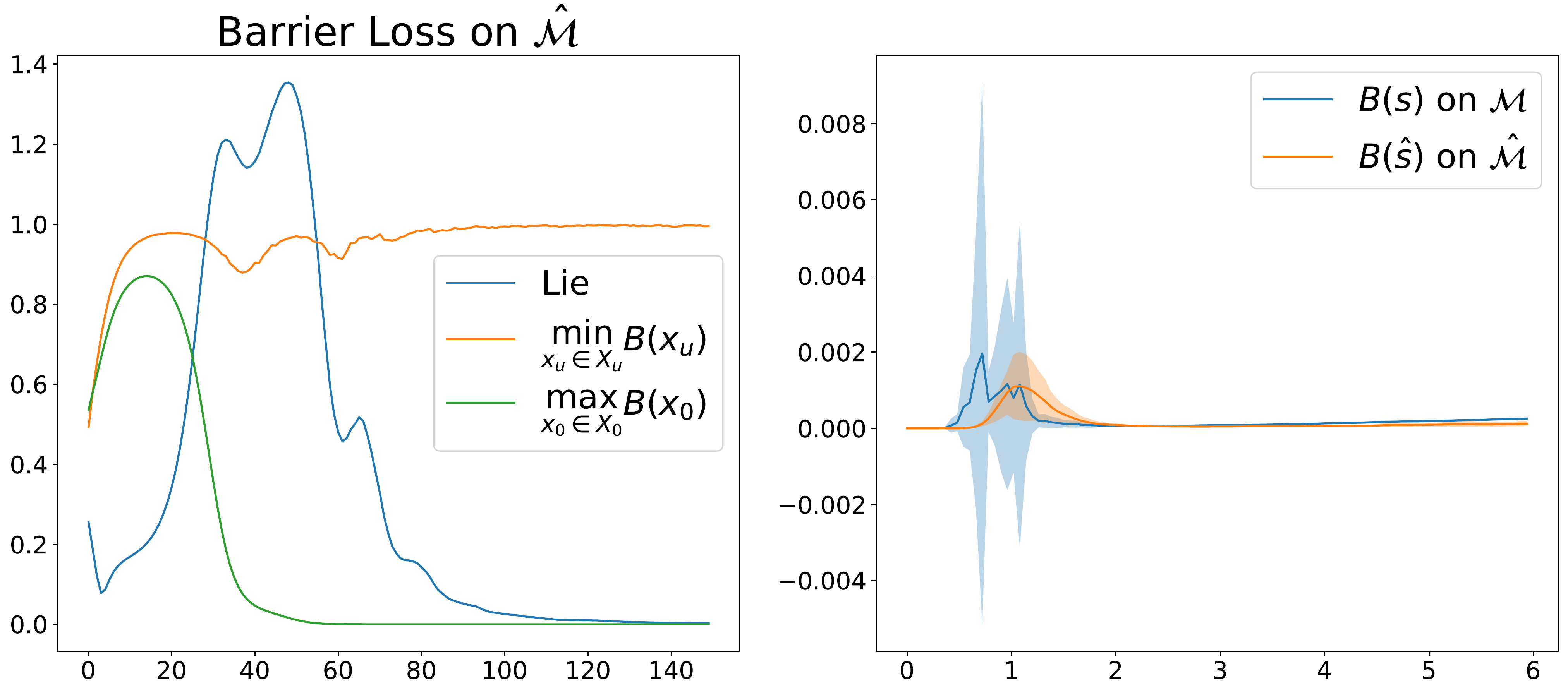}
    \caption{Barrier function training and testing in the 2D SDE example. The learned
barrier function maps the initial space to near 0 and the unsafe space to 1 with the third sample mean in Equation~\eqref{eq:barrier_loss}
to 0 (marked as Lie). The barrier function has a
similar close-to-0 value keeping constant along with the trajectories in the real environment and the generative model.}
\end{figure}
\begin{figure}
    \centering
    \includegraphics[width=\linewidth]{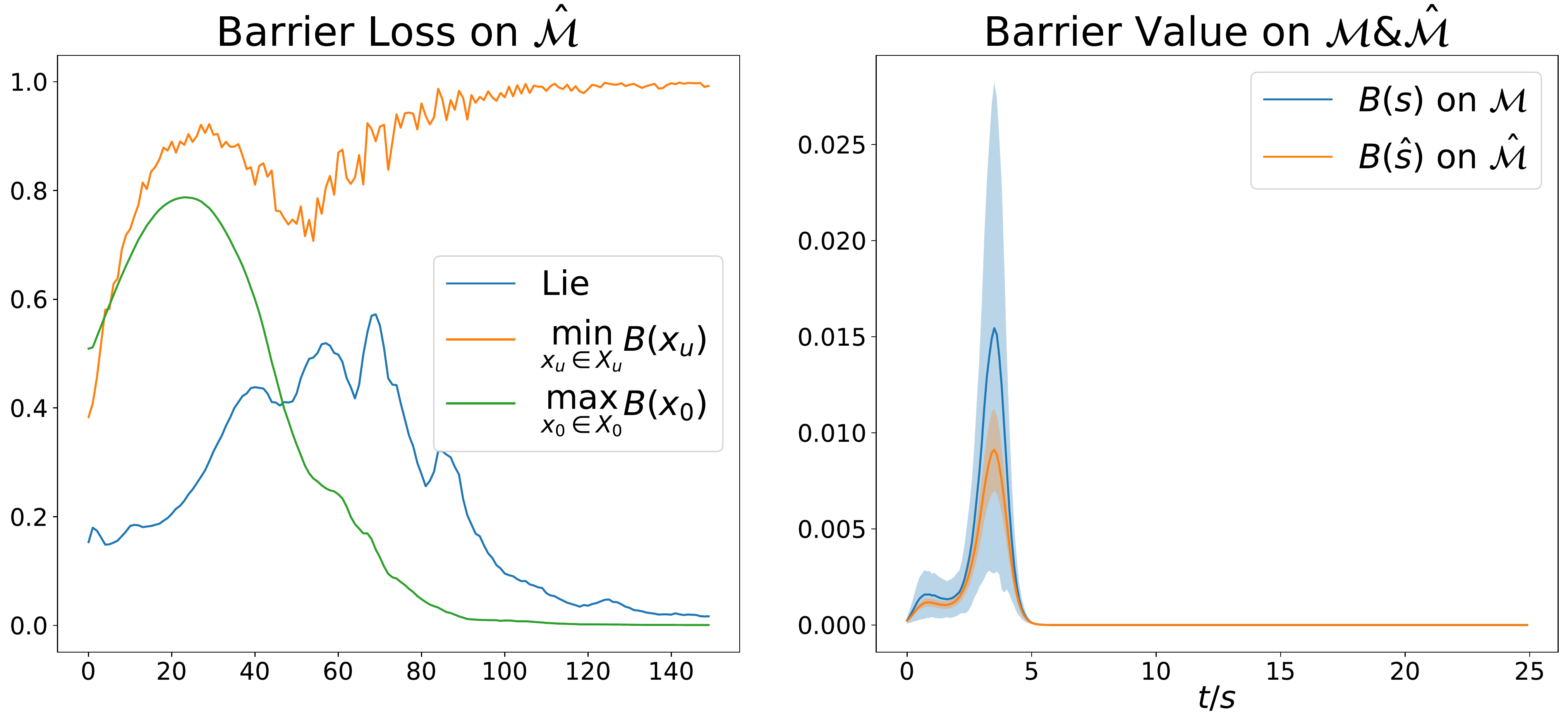}
    \caption{Barrier function training and testing in the Cartpole balancing example.}
\end{figure}

 \begin{figure}
     \centering
     \includegraphics[width=\linewidth]{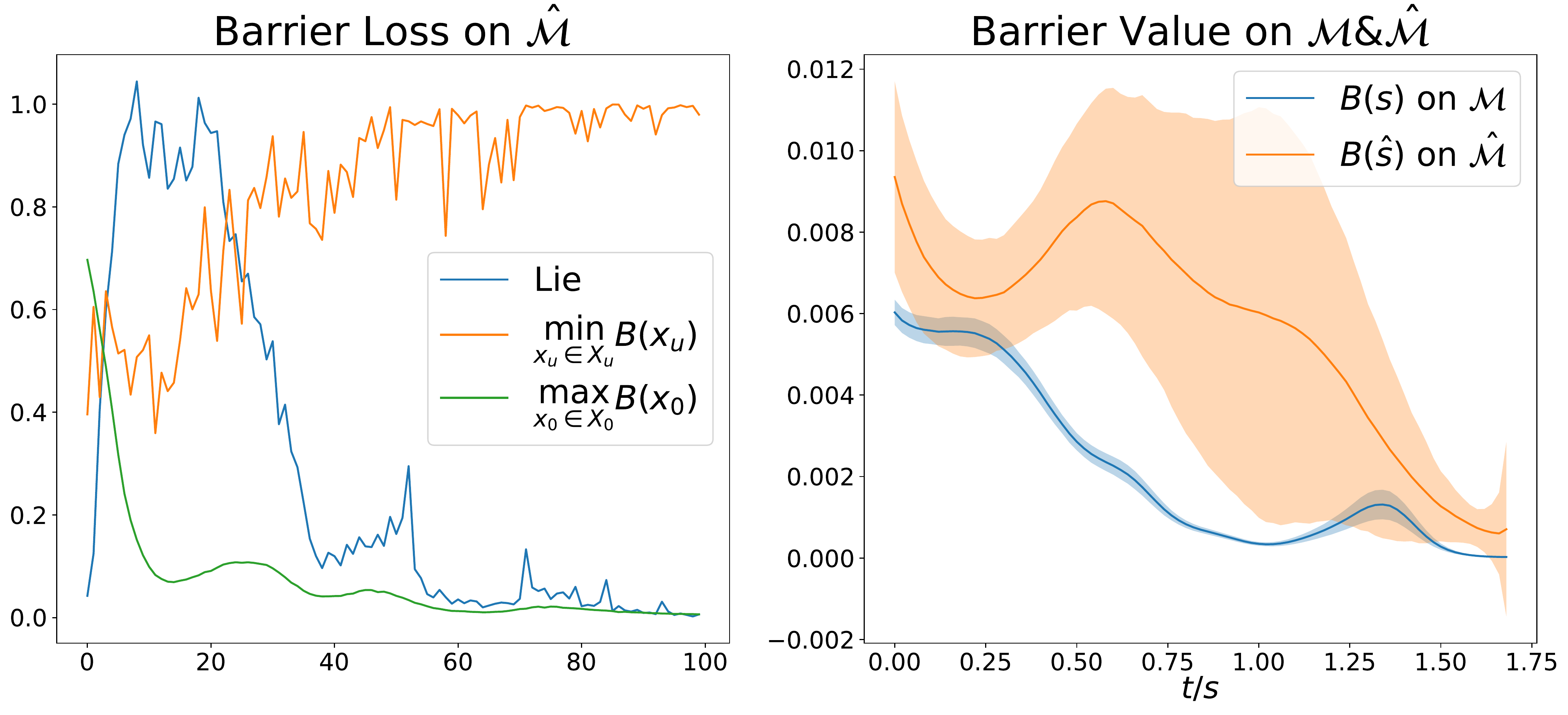}
     \caption{Barrier function training and testing in the UAV maneuvering example.}
\end{figure}

\begin{figure}
    \centering
    \includegraphics[width=\linewidth]{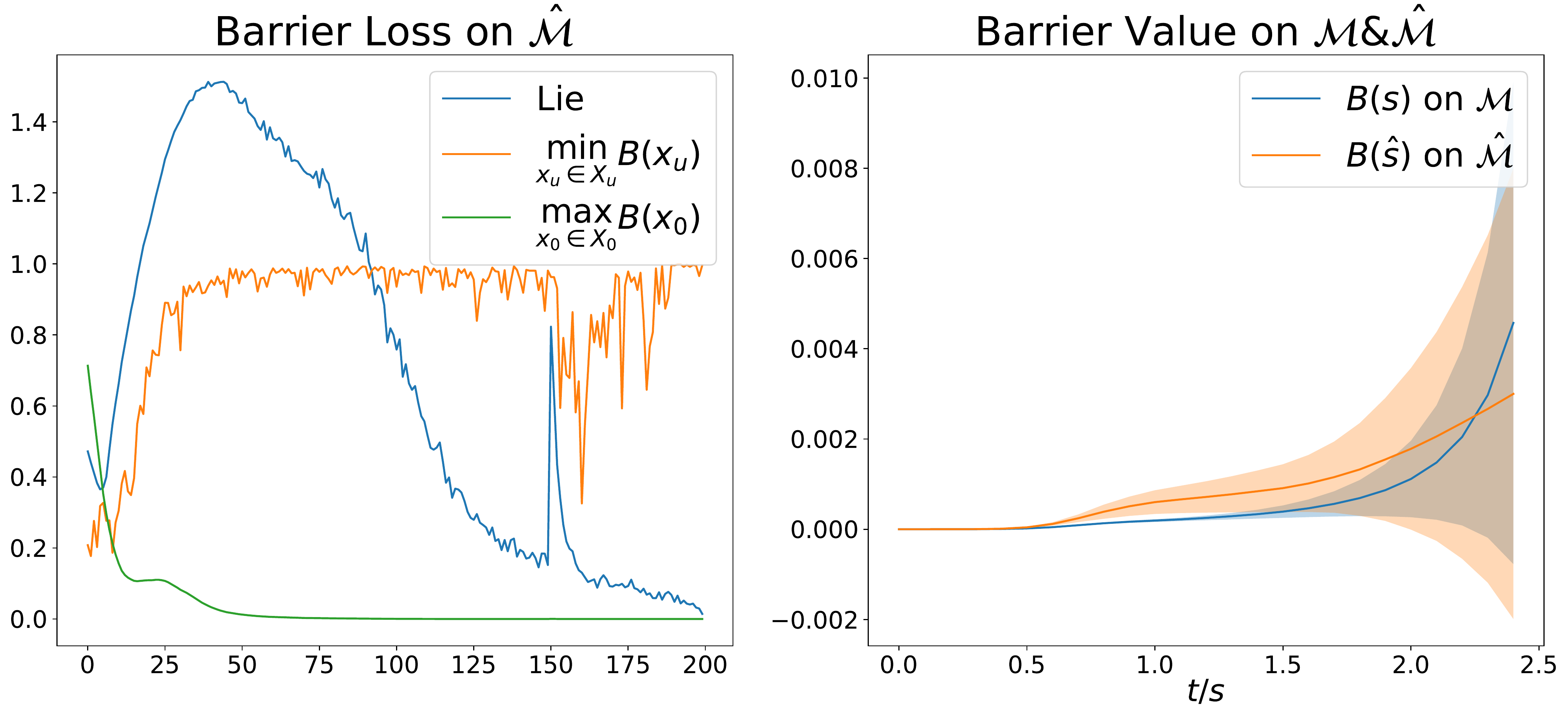}
    \caption{Barrier function training and testing in the Rocket powered landing example.}
\end{figure}

\yixuan{\textbf{Limitations:} As stated earlier, one key assumption of this work is the smoothness and continuity of the system behavior, which prevents its application to hybrid dynamics with jump conditions such as the contact dynamics in Mojuco and Safety Gym. One possible solution is to learn an ensemble generative model as a hybrid system to deal with those discontinuous contact dynamics, and we plan to explore it in future work. %
Another limitation of our framework is the computation complexity of the generative model (e.g., it takes around 8 hours to learn a policy for the Cartpole example and 1 day for the UAV and Rocket examples). In future work, we plan to improve the efficiency of this part by exploring techniques such as Continuous Latent Process Flows (CLPF)~\cite{deng2021continuous}.}
\section{Conclusion}\label{sec:conclusion}
We present a safe RL approach in unknown continuous stochastic environments that enforces hard reachability-based safety constraints through generative-model-based soft barrier functions. Our approach formulates a novel bi-level optimization problem and develops a safety RL algorithm that jointly learns the generative model, soft barrier function, and policy optimization. Experiments demonstrate that our approach can significantly improve empirical system safe rates over CMDP-based baselines and also provide a practical lower bound of safety probability.


\bibliography{example_paper}
\bibliographystyle{icml2023}

\end{document}